\documentclass[12pt]{article}
\usepackage{graphicx}
\usepackage{amsmath}
\usepackage{amssymb}
\usepackage[left=1in,top=1in,right=1in,bottom=1in]{geometry}
\usepackage{natbib}
\usepackage{setspace}
\makeatletter
\renewcommand\section{\@startsection{section}{1}{\z@}{-3.25ex plus -1ex minus -.2ex}{1.5ex plus .2ex}{\normalsize\bf}}
\renewcommand\subsection{\@startsection{subsection}{2}{\z@}{-3.25ex plus -1ex minus -.2ex}{1.5ex plus .2ex}{\normalsize\bf}}
\renewcommand\subsubsection{\@startsection{subsubsection}{3}{\z@}{-3.25ex plus -1ex minus -.2ex}{1.5ex plus .2ex}{\normalsize\bf}}
\makeatother

\newtheorem{thm}{Theorem}[section]

\newtheorem{prop}[thm]{Proposition}
\newtheorem{defn}[thm]{Definition}

\numberwithin{equation}{section}

\begin{document}
\begin{center}
\textbf{On (Some) Explanations in Physics}\footnote{I am indebted to Kyle Stanford, Dave Baker, John Manchak, Erik Curiel, and David Malament for helpful remarks on earlier drafts of this paper.  Thank you, also, to the Southern California Philosophy of Physics group, in particular Jeff Barrett, Craig Callender, and Tarun Menon for a spirited and productive discussion of the material presented here. I am also grateful to participants in the UC Irvine WIP seminar, especially Bennett McNulty, Sam Fletcher, and Cailin O'Connor, for their comments.  Finally, this paper benefited from the helpful remarks of two anonymous referees.}\\[1\baselineskip]
James Owen Weatherall\footnote{weatherj@uci.edu} \\Logic and Philosophy of Science \\
University of California, Irvine\\[1\baselineskip]
\end{center}
\singlespacing
\begin{center}\textbf{Abstract}\end{center}
I offer one possible explanation of why inertial and gravitational mass are equal in Newtonian gravitation.  I then argue that this is an example of a kind of explanation that is not captured by standard philosophical accounts of scientific explanation.  Moreover, this form of explanation is particularly important, at least in physics, because demands for this kind of explanation are used to motivate and shape research into the next generation of physical theories. I suggest that explanations of the sort I describe reveal something important about one way in which physical theories can be related diachronically.\\
\rule{6.5in}{1pt}\\[1\baselineskip]

\singlespacing
\begin{center}``What do we mean here by `explanation'? ... This whole issue, which perhaps lies between nature and sociology, seems to be a bit vague.  Quite possibly, an attempt to make the word \emph{explanation} more precise may do more harm to the field [of physics] than good.''\\ -Robert \citet[pg. 63]{Geroch}\\[2\baselineskip]\end{center}
\doublespacing

\section{Introduction}\label{introduction}

Consider the following questions, any of which might be heard in the halls of a physics department.
\begin{enumerate}\singlespacing
\item Our best theory of particle physics predicts that in very high energy experiments, which probe the smallest distance scales, the electromagnetic, weak, and strong forces should have approximately the same strength.  But at these same distance scales, gravitation is many orders of magnitude weaker.  Why is gravity so much weaker than any of the other forces?\label{naturalness}
\item The Standard Model of particle physics makes predictions that have been confirmed to 15 significant digits \citep{Odom+etal}.  But the Standard Model's predictions rely on 19 parameters that are ``put in by hand'' to agree with experiment; in order for the Standard Model to make accurate predictions at all, these parameters must be finely tuned.  Why do these Standard Model parameters take the values they do, and is there a sense in which they are determined by some underlying mathematical or physical principle?\label{fineTuning}
\item In Newtonian physics,\footnote{I will use the expression ``Newtonian theory'' interchangeably with ``Newtonian physics.''  In both cases I mean Newtonian dynamics plus gravitation.} inertial mass (the value $m$ that appears in $\mathbf{F}=m\mathbf{a}$) always has the same value as gravitational mass (the coupling to the gravitational field, i.e., the value $m$ that appears in $U_{\mathcal{G}}=m\varphi_{\mathcal{G}}$, where $\varphi_{\mathcal{G}}$ is the gravitational potential and $U_{\mathcal{G}}$ is the potential energy of a particle with mass $m$), even though in principle the theory distinguishes these masses.  This equivalence is empirical: it was first established (in slightly different terms) by Galileo; at the end of the 19th century, it was tested with very high precision by Lor\'and E\"{o}tv\"{o}s.  Yet the correspondence seems highly suggestive.  Why are inertial and gravitational mass equal in Newtonian physics?\label{correspondence}
\end{enumerate}\doublespacing
I need not multiply examples.  Each of these is a why question asked in a particular scientific context (physics).  As such, I take it that they are calls for scientific explanation.\footnote{Perhaps not all calls for explanation take the form of why questions, and perhaps not all why questions call for explanations.  But I claim these why questions do call for explanations.}  Indeed, they are why questions of a particularly important sort: these are the kind of questions that physicists often use to motivate their research projects.  Questions \ref{naturalness} and \ref{fineTuning} are open and form the basis of several major contemporary research programs\footnote{I do not mean ``research program'' in a technical philosophical sense.  I just mean that string theorists, loop quantum gravity theorists, and many particle phenomenologists (in the physicist's sense of phenomenology) take these questions to be central to their research.} in high energy particle physics and quantum gravity.  Question \ref{correspondence}, meanwhile, has been settled, or at least, we now have the theoretical machinery available to provide one sort of answer to it.  I claim that the answer one can now give in response to question \ref{correspondence} is an example of one kind of explanation that would satisfy the physicists who ask questions \ref{naturalness} and \ref{fineTuning}.  It may not be the only kind of explanation that physicists would ultimately deem satisfactory, though I think it is an ideal of the sort of explanation physicists have in mind when they ask questions \ref{naturalness} and \ref{fineTuning}.  My central goal in this paper will be to examine just what kind of explanation it is.

Before proceeding, however, I should give some context to the present discussion.  Over the last 20 years, since \citet{Salmon} proposed a detente between the causal and unificationist accounts of explanation, the idea that some sort of pluralist account of explanation is necessary to capture the full variety of explanatory phenomena has gained considerable support.  Salmon's own line was that the causal and unificationist accounts are not inconsistent.  Instead, he thought that they offer different kinds of understanding, corresponding to the different kinds of explanation.  On his view \emph{both} the causal and unificationist accounts are correct: the two accounts together offer a full account of explanation on which unificationist explanations are ``top-down'' and causal explanations are ``bottom-up.''  Any given event or phenomenon can be explained in both ways.  But this form of pluralism is still too limiting. It now seems that some explanations do not fit neatly into either account (see, for instance, \citet{Batterman}), and moreover, that some phenomena that are easily explained using one kind of explanation do not, as Salmon suggests, have explanations of the other sort.\footnote{I take \citet{Fisher}'s explanation of sex ratio in humans as an example of this latter sort.  (See also \citet[Ch. 1]{Skyrms}, where Fisher's work is put in perspective.)  A causal explanation can explain why any individual turned out to have a particular gender, but it cannot explain why the ratio must be what it is.}

More recently, \citet{Godfrey-Smith} (following \citet{Kuhn}) has suggested a different kind of pluralism, in which what counts as a good explanation can vary depending on scientific context.  Explanation in biology need not be the same as explanation in physics, and explanation in either field in the early 21st century need not be the same as explanation was in, say, the 17th century.  On this view, it is a mistake (as Geroch suggests above) to attempt to characterize scientific explanation in advance: what counts as a good explanation in science is evolving along with the sciences themselves.  I am very sympathetic to this view.  But I take it that two kinds of project remain, even after a pluralistic, contexualist account has been accepted.  The first kind of project is to identify the working parts of contextualist pluralism.  Certainly, scientific context determines the array of explanations that are available.  But in many scientific contexts, such as contemporary physics, it seems that more than one form of explanation is common.  Moreover, as Salmon suggests, many phenomena may have explanations of radically different types.  What makes a particular explanation an appropriate one in a given instance, as an answer to a particular why question?\footnote{I would suggest that the problem of identifying appropriate explanations is essentially pragmatic, but will defer discussion of this point to future work.}

The second kind of project, meanwhile, is to identify interesting explanations used in various contexts and attempt to understand their epistemic and scientific virtues.  The current paper is an example of this second kind of project.  I should say, however, that once one has accepted some form of pluralism, merely taxonomizing the possible forms of explanation is not fruitful.  Nevertheless, some examples of explanation stand out as particularly important for the practice of science.  With this in mind, let me show my hand.  I will presently argue that the explanation I describe in the next section of this paper is not well treated by the standard philosophical accounts of scientific explanation.  This argument will form the main thrust of the paper, but by itself the thesis need not be remarkable.  Instead, I take the principal moral of the present discussion to be that the explanation I will describe, taken in conjunction with the many demands for such explanations that seem to arise in the practice of contemporary physics (as witnessed by the two other questions I mention above), reveals something important about one way in which physical theories may be related diachronically.  I will suggest that the expectation that certain cross-theoretic explanatory demands will be met by future theories is an essential part of inquiry in physics that is sometimes obscured by the traditional accounts of scientific explanation.

From here I will proceed as follows.  I will start by clarifying what question \ref{correspondence} is asking.  I will then sketch what I take the answer to be.\footnote{The discussion in the body of the paper is precise, but informal.  The technical details of the explanation are included in an appendix.}  This question has been answered informally in a variety of ways since General Relativity (GR) first appeared.  The answer I will present here is certainly in the spirit of these standard responses, though it precisifies a number of details about the relationship between mass in GR and Newtonian physics that are usually left vague.  To my knowledge, the form of the answer I will present here is original and may be of independent interest.  After presenting the explanation I have in mind, I will turn to the question of whether the explanation I offer here can be understood within the rubrics of various well established accounts of explanation.  I will conclude that it cannot.  In the remainder of the paper, I will try to articulate how the present explanation works, highlighting its distinctive features and arguing for the moral suggested above.

\section{Why are inertial and gravitational mass equal in Newtonian gravitation?}\label{explanation}

As I have said, inertial mass and gravitational mass are conceptually distinct in Newtonian physics (I will distinguish them here by writing $m_{\mathcal{I}}$ for inertial mass and $m_{\mathcal{G}}$ for gravitational mass).  Indeed, one would expect them to be unrelated to one another.  Inertial mass is a constant of proportionality in the fundamental dynamical principles of the theory.  It appears in Newton's second law, which states that $\mathbf{F}=m_{\mathcal{I}} \mathbf{a}$; momentum is defined as $\mathbf{p}=m_{\mathcal{I}} \mathbf{v}$; kinetic energy is $T= 1/2 m_{\mathcal{I}} v^2$.  One can think of inertial mass as a measure of a body's tendency to accelerate under the influence of an impressed force.  Inertial mass is closely related to inertial \emph{motion}, which enters Newtonian theory via Newton's first law.  The first law states that a body undergoing uniform rectilinear motion will not deviate from that motion unless acted on by an external force; inertial mass, then, determines a body's tendency to deviate from uniform rectilinear motion when acted on by an external force, whether gravitational or otherwise.

Gravitational mass, meanwhile, determines the strength of the gravitational force that a body experiences in a gravitational field.\footnote{A distinction is sometimes made between ``passive gravitational mass'' and ``active gravitational mass''.  Passive gravitational mass is what I have described above; active gravitational mass, meanwhile, can be understood as a measure of the strength of the gravitational field produced by a body.  The status of active gravitational mass in geometrized gravitational theories is of some independent interest, but in the present paper I am exclusively interested in passive gravitational mass.  Whenever I use the expression ``gravitational mass,'' the reader should assume I mean ``passive gravitational mass.''}   If a test body\footnote{By test body, I mean a body that is assumed not to contribute to the gravitational field itself.  In other words, when considering test bodies one neglects the ``backreaction'' of a body's own gravitational field.} with gravitational mass $m_{\mathcal{G}}$ is placed in a gravitational potential $\varphi_{\mathcal{G}}$, then the body will have gravitational potential energy $U_{\mathcal{G}}=m_{\mathcal{G}}\varphi_{\mathcal{G}}$ and will experience a force $F_{\mathcal{G}}=-m_{\mathcal{G}}\nabla\varphi_{\mathcal{G}}$.  Gravitational mass can be thought of as gravitational \emph{charge}, in analogy with classical electric charge.  Indeed, the fundamental force equations have exactly the same structure.  A test charge $q$ in an electric potential $\varphi_{\mathcal{E}}$ will have electrical potential energy $U_{\mathcal{E}}=q\varphi_{\mathcal{E}}$ and will experience a force of $F_{\mathcal{E}}=-q\nabla\varphi_{\mathcal{E}}$.

The parallel with electric force is particularly salient here.  Suppose one wants to know the acceleration exhibited by a test particle of charge $q$ and inertial mass $m_{\mathcal{I}}$ in an electric potential $\varphi_{\mathcal{E}}$.  Combining Newton's second law with the force law for a test particle in an electromagnetic field, one finds that\[\mathbf{a}=-\frac{q}{m_{\mathcal{I}}}\nabla\varphi_{\mathcal{E}}.\]  In other words, the acceleration depends on the ratio of the charge to the inertial mass of the body, both of which are freely varying, independent quantities.  One can find in nature bodies with many different values for the ratio $q/m_{\mathcal{I}}$.  Meanwhile, if one performs the identical calculation to determine the acceleration due to gravity (given a fixed gravitational potential), one likewise finds,\begin{equation}\label{standardAcceleration}\mathbf{a}=-\frac{m_{\mathcal{G}}}{m_{\mathcal{I}}}\nabla\varphi_{\mathcal{G}}.\end{equation}  Again, the acceleration depends on the ratio of two values: the gravitational and inertial masses.  Given the structural similarities between the gravitational and electric cases, one should expect to go out into the world and find bodies with a wide array of different values for the ratio $m_{\mathcal{G}}/m_{\mathcal{I}}$.  After all, (a) how much a body will tend to deviate from rectilinear motion given an external force and (b) the strength of that external force should be independent quantities.  But when we start looking into how bodies behave in a gravitational potential, we find something quite different.  Given any body at all, the ratio $m_{\mathcal{G}}/m_{\mathcal{I}}$ always takes the same value: choosing the natural units, we always find that $m_{\mathcal{G}}/m_{\mathcal{I}}=1$.

Given this background, the \emph{explanandum} can be stated as follows: all evidence suggests that, given any body, the gravitational and inertial masses of that body are equal, despite the fact that Newtonian theory gives no reason to expect these two masses to be related.  We want to understand why this is.  In some ways it is an unusual why question (at least with respect to standard accounts of explanation), so I want to spend some time up front focusing on its distinctive features.  First off, it is a question about a general observational feature of the world, but it is expressed in the terms of a specific physical theory.  In other words, the question takes the Newtonian concepts of gravitational and inertial mass for granted.  One can express the observational fact without reference to the Newtonian theory---Galileo first described the phenomenon that all bodies fall at the same rate, irrespective of mass, before Newton was born---but when one does so, the question does not arise.  One might, perhaps, wonder about the apparent universality of free fall in other contexts or even quite generally: after all, Galileo's results conflicted with the Aristotelian expectation, so one might well have demanded an explanation for Galileo's observations in the context of \emph{Aristotelian} physics, too.  But without the background conceptual machinery of Newtonian physics, the question is different.  I am interested in a specific question \emph{about} the world, expressed \emph{within} the Newtonian framework.  It seems to me that questions \ref{naturalness} and \ref{fineTuning} are similar in this regard, \emph{mutatis mutandis}.

This first feature suggests a second feature.  Although the question is posed within the Newtonian framework, and cannot be quite the same question if posed in other contexts, it cannot be answered within the Newtonian framework.  As we have seen, the Newtonian concepts of gravitational and inertial mass are distinct.  That the two masses always take the same value for any given body is contingent, though suggestive, within Newtonian physics. But this means that any appropriate answer will have to go beyond Newtonian physics.  So we have a question posed in one theory that can only be adequately answered by appealing to another, presumably more general or fundamental, theory.

But what form could such an answer take?  \citet{Einstein} claimed that the observed equivalence between gravitational and inertial mass was an important factor in his development of GR, and that it can be taken as evidence in favor of the ``general postulate of relativity.''  The principle that the two mass concepts should be identified, or equivalently the idea that free fall does not depend on mass, is often called the weak equivalence principle and continues to play a central role in some presentations of GR (see, for instance, \citet[ch. 1]{Weinberg}).  From this point of view, the observed equivalence is explained by asserting that in a supervening theory (GR), no distinction is made between the two masses.  Inertial and gravitational masses are simply the same.  But there is something strange, and potentially misleading, about this answer to the original question.  The reason for the difficulty is that a more accurate account of the situation in GR, using only concepts native to GR itself, would be to say that there is \emph{only} inertial mass.\footnote{And even inertial mass is problematic except in the case of a test body, unless a spacetime satisfies very strong symmetry conditions.}  It is not that gravitational mass is explicitly identified with inertial mass, but rather than gravitational mass has been stricken from the theory altogether.  There is no gravitational potential in GR, nor is there a gravitational force, and so there is no parameter relating gravitational force to a background gravitational field.\footnote{This point is put clearly, for instance, by \citet{Sachs}, though it has not always been recognized by the physics community, as \citet{Weinberg} attests.}  When one attempts to answer the original question by appealing to some supposed equivalence between the two kinds of mass in GR, one mixes terms from two theories in a way that is dubious and confusing.

These considerations suggest another response to our original question.  Given that gravitational mass does not make sense in GR, one might say that the question turns out to be an error.   We used to think that gravitation was a force (one might say), and that a body's gravitational mass determined the magnitude of the gravitational force exerted on the body in a gravitational field.  But now we know that gravitation is not a force at all and so questions about gravitational mass do not make any sense.  This type of response is intended to dissolve the question, by directing the questioner to a textbook on GR.  But I claim that this kind of response is unsatisfying.  First, the question was asked in a specific framework; to say that that framework is no longer widely accepted is irrelevant.  Second, even if one accepts that GR supercedes Newtonian physics, and one accepts moreover that gravitational mass does not exist in GR, an important question remains.  In Newtonian physics, which everyone accepts as a predictively accurate theory in some regimes, there are two different kinds of mass.  As such, I can point to two kinds of roles that one might expect mass to play.  Supposing that GR is correct that gravitational mass does not make sense, why is the world so well-described, at least in certain regimes, by a theory that supports two concepts of mass?\footnote{Curiously, one can already imagine giving the same response to question \ref{naturalness}: why should we compare gravitation to the other three forces?  After all, we already know that there is no such thing as gravitational force!  But this answer would be equally unsatisfying in that context, for the same reason: gravitation \emph{is} conceived of as a force in modern particle physics, which is at least part of the difficulty in making quantum field theory and GR mesh.  I think this serves to underscore the curious character of the questions above: they are all expressed in the language of one theory, but one fully expects that the answer will come from a theory in which the terms of the question may not make any sense.}

I think this discussion helps to clarify both what the question really is and what kind of answer would be appropriate.  We have two fixed points to navigate between.  First, the question is such that it cannot be answered by the Newtonian theory.  New physics is required.  But second, it is a question that needs to be answered in the terms in which it was asked, i.e., within the Newtonian framework.  As we have seen, using Newtonian terms within the framework of GR leads to inconsistencies and serious confusion, while using concepts native to GR at best allows a dissolutive response, rather than an answer, to the question.  With these two points in mind, one might rephrase the question once again, as follows.  Given that we now believe GR to have superseded Newtonian gravitation as our best theory of large-scale dynamics and gravitation,\footnote{Some readers might balk at this point and argue that GR cannot provide \emph{any} kind of explanations whatsoever if it is not true.  I do not share the intuition that only true theories have explanatory power.  But I do not think this issue is relevant here. Everything I say in what follows can be recast in conditional form: explanations of the form ``If GR were true, then...'' are perfectly sufficient for a discussion of what kinds of explanation are possible in science.} why are gravitational and inertial mass equal in Newtonian theory?\footnote{Rephrasing the question in this way is only possible when one can point to the superseding theory.  In the cases of questions \ref{naturalness} and \ref{fineTuning}, no superseding theory is known.  These questions might be rephrased in terms of a future possible theory, or as a statement of a certain kind of research objective.  One is looking for a new theory $T$ that can tell us why, given that theory $T$ supersedes the Standard Model (say), the parameters in the Standard Model take the particular values that they are observed to take.}  This phrasing presupposes that even though GR has virtues that Newtonian gravitation lacks, there are still regimes in which Newtonian gravitation provides a satisfactory characterization of nature.

To answer the question, one needs to show, in detail, how Newtonian theory relates to GR.  One needs to show why, if GR is true, Newtonian theory is such an effective theory in some contexts.  One way of doing this would be to show that Newtonian theory can be reached from GR in an appropriate limit that captures the circumstances in which Newtonian theory seems so effective.\footnote{Many philosophers have questioned when and whether it is possible to show that an old theory reduces to a new theory (in the philosopher's parlance) or a new theory reduces to an old theory (in the language of working physicists).  \citet{Nagel1961, Nagel1970}, for instance, treated intertheoretic reduction as explanation (in the deductive-nomological sense) in a way that bears a rough family resemblance to what I am thinking of; \citet{Nickles}, meanwhile, argues that often explanatory reductions are not possible at all (at least in the DN sense).  Curiously, Nickles identifies Newtonian physics and relativity theory as a prime example of a reduction relationship that is \emph{not} an explanation in the DN sense, but rather a collection of rough intertheoretic relations.  It seems to me that if \emph{any} intertheoretic relationship deserves to be called deductive, it is the relation between Newtonian theory and GR.  But I do not intend to enter a debate on intertheoretic reduction here.  Instead, I want to distinguish \emph{identifying} reduction and explanation (as Nagel does) from a more ambiguous demand that a new theory explain, at least in some sense, why our old theories succeeded.  One way of cashing this requirement out is to say that a new theory cannot make predictions that are inconsistent with the experimental successes of previous theories (in which case the new theory will at least reveal regularities captured by the old theory).}  This observation suggests yet another refinement of the original question.  What we are really interested in is the following: Given that we now believe GR to have superseded Newtonian theory, then why, \emph{insofar as Newtonian theory is a limiting case of GR}, are inertial and gravitational mass equal in Newtonian theory?  I maintain that this reformulation captures of the spirit in which the original question was asked, and moreover, we finally have a question that is precise enough to answer.  It is now clear that the answer would involve trying to show that if Newtonian theory is taken as the limit of GR in the appropriate way, then Newtonian theory \emph{does} support two concepts of mass, and moreover, that for any body these two masses must be equal.  In other words, one shows that gravitational mass arises in some way in the limiting process, and that the result must be equal to inertial mass.\footnote{\label{reformulation}At the very least, the answer sketched would be responsive to the question when reformulated as relative to GR.  It seems to me that, given GR and the relationship between GR and Newtonian physics, the correct way of thinking about Newtonian physics is through the lens of GR.  But not all philosophers agree: there is another way of thinking according to which Newtonian physics is an entirely unrelated (and false) theory of spacetime and gravitation.  All I can say in response is that I agree that if one understands Newtonian physics in this way, then the explanation I offer is not satisfying.  But I disagree that this is the most natural way of understanding Newtonian physics.  In particular, it seems to me that the relationship that one \emph{does} get between Newtonian physics and GR by understanding the former as a limiting case of the latter is highly desirable from the point of view of practicing physics (and philosophy of physics).}

It turns out that it is possible to perform this procedure exactly as described.  (See appendix.)  There is a precise sense in which Newtonian theory arises as a limit of GR \citep{Kunzle,Ehlers,MalamentNG}.\footnote{\label{motivation}This footnote is in some sense a continuation of footnote \ref{reformulation}; it also addresses a possible worry about so-called ``counterfactual'' explanations.  There is a way of thinking about Newtonian gravitation in light of GR according to which describing the sense in which Newtonian theory is a limit of GR is tantamount to describing the experimental conditions under which it is appropriate to use Newtonian laws for purposes of approximation.  This is not what I have in mind.  When I describe Newtonian theory as a limiting case of GR, I mean that given a fixed background manifold, (a) there is a precise sense in which classical spacetime structure on that manifold can be understood as the limit of a sequence of relativistic spacetime structures, and (b) when one takes this limit (with appropriate background assumptions), one recovers the correct Newtonian laws \emph{in toto} on the resulting classical spacetime.  (See appendix.)  There is at least one place in which the two ways of thinking make contact, however.  If, rather than considering experimental settings in which particular Newtonian laws are appropriate approximations, one instead considers the experimental circumstances under which one would be unable to identify failures of absolute simultaneity (perhaps because the apparatus is too coarse grained to detect the finite speed of light), then it is appropriate (from an experimental point of view) to assume that spacetime has a classical, Galilean structure.  And if one were to make such an assumption, it would then be appropriate to take the limit I have described to derive the relevant approximate laws---hence ``motivating'' the explanatory power of the limit.  I also take it that describing the limit in this way should moderate worries that the limit ``as $c$ goes to infinity'' cannot be explanatory because it is counterfactual, since one can understand the limit as describing the case where for all we can tell $c$ \emph{is} infinite (as was the situation in 1687).}   It involves a two step process.  One begins by considering a one-parameter family of relativistic spacetimes, parametrized by some variable $\lambda$.  $\lambda$, at the present level of discussion, can be taken to reflect the inverse of the ``speed of light''\footnote{There is an abuse of language, here.  Really, $\lambda$ parametrizes something about the metrical structure of the spacetime, specifically how wide the light cones are at a point.  But the widths of the light cones indicates how null vectors, which are the possible tangent vectors for the wordlines traversed by light, relate to the timelike and spacelike vectors in the spacetime.  Hence wider lightcones indicate ``faster'' light.} in each of the spacetimes.  By constructing this family of spacetimes carefully, one can consider the limit that the spacetimes approach as $\lambda$ approaches 0 (corresponding to taking the speed of light to infinity).  The result is a degenerate ``classical'' spacetime with many of the characteristic features of Newtonian physics: space is always flat and Euclidean; there is a unique sense of space at a time and absolute simultaneity holds; the spacetime is Galilean relativistic, which means that (1) measurements of elapsed time and the distance between simultaneous events will be the same for all observers, irrespective of their motion and (2) there is no absolute standard of rest. But we have not yet recovered Newtonian physics.  Rather, we have reached an intermediate point between GR and Newtonian physics, which I will call geometrized Newtonian theory.\footnote{Geometrized Newtonian theory is sometimes called Newton-Cartan theory. It was first developed during a lecture course by \'Elie Cartan in the early 1920s, as an attempt to understand how Newtonian gravitation related to GR \citep{Cartan1, Cartan2}. (See also \citet{Friedrichs}.)}$^, $\footnote{It may be helpful to emphasize that geometrized Newtonian theory is \emph{not} a model within GR that is somehow suggestive of classical physics.  It is an independent theory with the empirical content of Newtonian physics.  Indeed, a classical spacetime as described here is not and could not be a relativistic spacetime because it does not have an appropriate metric structure.}  In geometrized Newtonian theory, gravitation is still geometrical: the geometry of spacetime is curved, with curvature determined by the distribution of matter in the universe, and gravitational effects are manifestations of the resulting geometry.  Importantly, since gravitation is geometrical rather than a force between bodies, gravitational mass does not make any more sense in the context of geometrized Newtonian theory than in GR.

To see where gravitational mass comes from, we need to take the second step in the limiting process.  This step makes use of a theorem due to Andrzej Trautman \citep[Prop. 4.2.5]{MalamentGR}.  Trautman's theorem tells us that, given a classical spacetime of the sort found in geometrized Newtonian theory, satisfying certain conditions, one can find\footnote{See the appendix for details of this claim, and for a precise statement of the conditions a (geometrized) classical spacetime needs to satisfy in order to recover standard Newtonian theory, which are not trivial.} (1) another spacetime that is flat,\footnote{A classical spacetime as considered by geometrized Newtonian theory is generally curved, though \emph{space} is always flat.  In standard Newtonian physics, spacetime taken as a whole has to be flat.} and (2) a scalar field $\varphi_{\mathcal{G}}$ that satisfies Poisson's equation (the relationship between the distribution of matter in the universe and the gravitational potential in Newtonian gravitation) and which is such that for any free (inertial) massive test point particle, \begin{equation}\label{acceleration}\mathbf{a}=-\nabla\varphi_{\mathcal{G}}.\end{equation}  Thus, under certain circumstances, we can recover a flat spacetime and a gravitational potential $\varphi_{\mathcal{G}}$ that has just the relations to both the distribution of matter in the universe and the dynamics of a particle that we would expect from Newtonian physics, and moreover, in this flat spacetime, the particle trajectories determined relative to the gravitational field agree with the particle trajectories as determined by the geometrized theory in the initial curved spacetime.  More loosely, one finds a flat spacetime and gravitational field that ``makes the same predictions'' as the geometrized theory.  We have now recovered full-blown, standard Newtonian physics as a limit of GR.

I want to draw attention to an important feature of Eq. \eqref{acceleration}.  It is a derived relation between the acceleration of a particle and the gradient of the gravitational potential.  If we use $\mathbf{F}=m_{\mathcal{I}}\mathbf{a}$, Eq. \eqref{acceleration} implies that the force on a massive test point particle arising from the gravitational potential $\varphi_{\mathcal{G}}$ is\begin{equation}\label{mi=mg}\mathbf{F}=m_{\mathcal{I}}\mathbf{a}=-m_{\mathcal{I}}\nabla \varphi_{\mathcal{G}}.\end{equation}  Eq. \eqref{mi=mg} tells us directly that the coupling to the gravitational field in Newtonian physics is given by the \emph{inertial} mass.  Thus the reason that gravitational and inertial mass are always equal is that gravitational mass simply is inertial mass.  And so we have an answer to the original question.

\section{What have I just done?}\label{surveyTypes}

Now that I have offered an answer to one of my questions, I can ask what kind of explanation I have given.  It seems clear without further argument that this explanation is \emph{not} a causal explanation, in any of the senses of causal explanation that have been articulated over the last few decades.\footnote{Some prominent examples are of course \citet{SalmonCE}, but also \citet{Cartwright} and \citet{Woodward}.  \citet{Strevens, StrevensD} also offers a kind of causal account of explanation, though his ``kairetic account'' also includes some of the desirable features of the unificationist account.}  It likewise does not fit into any of the earlier statistical accounts of explanation, such as the statistical relevance account of explanation or the inducto-statistical account.\footnote{For more on either of these, see  \citep{Salmon}.}  In the remainder of this section, I will focus on two other prominent accounts of explanation that, at least \emph{prima facie}, have a better chance of capturing the kind of explanation I gave in the previous section: the deductive-nomological (DN) account and the unificationist account.\footnote{\citet{Batterman} also discusses a form of explanation that is not well-treated by the causal and unificationist accounts.  He dubs it ``asymptotic explanation.''  I think that he has correctly identified a form of explanation that the received accounts miss; however, I want to emphasize that the present example is different from the examples Batterman offers.  Asymptotic explanation involves explanations of ``universality''---properties that families of systems have at a given distance scale, irrespective of their microscopic details.  The explanation of such universal features involves the renormalization group, which is a method for moving between different levels of description of a physical system.  The form of explanation described here relies on a very different kind of limiting procedure and concerns relations between theories, not between distance scales.
It may well turn out that the answers to questions \ref{naturalness} and \ref{fineTuning} described at the beginning of this paper \emph{will} turn out to look more like Batterman's examples.  (This is especially likely for question \ref{fineTuning}.)  But what is important is that there are cases in which the kind of explanation I am describing is decidedly \emph{not} asymptotic explanation.}

\subsection{The Deductive-Nomological Account}

The DN account of explanation, originally proposed by \citet{Hempel+Oppenheim}, was long the received view of explanation.  On this account, an explanation is a (logical, first-order) argument by which the thing to be explained, the \emph{explanandum}, is deduced from a set of true premises, the \emph{explanans}.  It is taken to be necessary that the \emph{explanans} include at least one law of nature; generally, it will also include particulars such as initial conditions or boundary conditions.  The thing to be explained is a proposition.  The intuition is that an explanation is a demonstration of law-like expectability.  To explain a proposition is to show that one could expect it to be true, given the laws of nature and some given set of circumstances.\footnote{See, as ever, \citet{Salmon}.  I am glossing over many difficulties concerning what might count as a law of nature and what kind of deduction is necessary.}

Does the explanation I give in section \ref{explanation} fit the DN mold?  It certainly has many of the central features of a DN explanation.  The explanation consists of an argument by which the \emph{explanandum} is derived.  The \emph{explanans} is perhaps a bit broad: all of the central principles of GR must be included in order to set up the limiting process necessary to recover geometrized Newtonian theory.  But among these principles are several law-like propositions.   Finally, there is a strong sense in which the argument's explanatory power comes from the fact that the result is derived from these central principles of GR, which means that the explanation is nomological in an important way.

But not all is well. For one, the technical details of the explanation I describe above make use of powerful mathematical tools.  It is not at all clear that first-order logic has the resources to express the full mathematical argument properly.  A second worry concerns the language in which the deduction is to be performed.  If one were to begin with a language consisting only of terms readily interpreted in GR, then gravitational mass would not appear and so the proposition ``inertial and gravitational mass are (always) equal'' could not even be expressed.  Conversely, one might begin with an augmented language that includes gravitational mass terms and then endeavor to show that gravitational mass is always equal to inertial mass via a deduction in this augmented language.  But it is difficult to see how to interpret such an augmented language.  In any case GR by itself would not provide a model for the language, which means that even if the deduction were to succeed, it would not follow that one had derived the \emph{explanandum} by appeal to GR alone.  Either way it seems that an essential part of the explanation cannot be captured within the DN framework.

So the present explanation bears some family resemblance to DN explanations, though it does not have just the form described by Hempel and Oppenheim (because it requires the full resources of mathematical physics).  More troubling, the DN account does not naturally permit the final (important) step of the explanation.  All that said, if one were committed to the claim that \emph{all} explanations must fit the DN model, it might be possible to extend the model to include the present example.  Alternatively, if disinclined to extend the DN model, one might bite the bullet and claim that the example does not have any explanatory power.  I do not want to make either of these moves, however.  Insofar as I already accept a pluralistic view about explanation, I do not see any virtues in a procrustean reading of the present explanation as an example of DN explanation.

There is another consideration here.  Many writers have argued, I think successfully, that Hempel and Oppenheim offer neither necessary nor sufficient conditions for an argument to be explanatory \citep{Godfrey-Smith, Salmon, Kitcher}.  In other words, even if the present explanation \emph{does} fit the DN model, it does not follow that it is explanatory \emph{because} it fits that model.  One still needs to give an account of what makes a given argument explanatory.  A prominent attempt to describe what additionally may be required to make an argument explanatory is given by the unificationist account of explanation, which I will turn to presently.

\subsection{The Unificationist Account}

\citet{Kitcher} provides the authoritative manifesto on the unificationist account of explanation, so I will focus on the version of the account given there.\footnote{For more on the history of the unificationist account, see \citet{Kitcher} and \citet{Salmon}, as well as references therein.}  The basic idea of the unificationist account is that science aims to explain phenomena by showing how a phenomenon fits into a unified systematization of our beliefs.  Kitcher makes this idea precise in the following way.  Suppose that $K$ is the set of all statements endorsed by the scientific community.  An acceptable explanation is a member of the \emph{explanatory store} over $K$, denoted $E(K)$.  How can we characterize the members of $E(K)$?  First, we say that $E(K)$ is a set of arguments by which some members of $K$ are derived from other members of $K$.  But in general there will be many such sets of arguments.  To pick one, Kitcher introduces \emph{general argument patterns}, ordered triples consisting of (1) a \emph{schematic argument}, which is a sequence of sentences with key terms replaced by dummy variables (i.e. a sequence of \emph{schematic sentences}); (2) a set of sets of filling instructions corresponding to the schematic argument, where a given set of filling instructions tells you how to fill in the dummy variables of a schematic sentence; and (3) a classification of the schematic argument, which tells you which of the sentences in the argument are supposed to count as premises and which are conclusions.  One general argument pattern is more \emph{stringent} than another if its classification and the structure of its schematic sentences together make it more difficult to instantiate.  Given this machinery, Kitcher says that $E(K)$ consists of the minimal set of maximally stringent general argument patterns from which a maximal number of conclusions can be drawn.  More roughly, one wants to find the smallest subset of $K$ from which the other members of $K$ can be derived, using the fewest possible stringent argument patterns.

As on the DN account, an explanation on the unificationist account is an argument.\footnote{Depending on how strictly one construes the constraints on what counts as an argument for the unificationist account, the two worries expressed with regard to the DN model may carry over.  I will proceed by assuming that the unificationist account allows some flexibility about what counts as a permissible argument.}  But the unificationist account offers an additional set of constraints on the kinds of arguments that count as explanatory, above and beyond the conditions of the DN account.  It is not sufficient (or necessary) that a given conclusion be derived from a set of premises including a law, as in the DN model; now, an argument is explanatory if it is a member of the explanatory store of a set of statements endorsed by the scientific community.\footnote{On Kitcher's account, one would add an ``only if'' to this last sentence.  But if we adopt a pluralist view of explanation on which some, but not necessarily all, explanations are unificationist explanations then we want to understand the condition of membership in $E(K)$, as defined by Kitcher, to be only a sufficient condition.}  To understand whether the explanation in section \ref{explanation} is a unificationist explanation, then, we want to determine whether we should expect it to be in the explanatory store over $K$. There are several reasons to think that it should not be.

First, it is not clear that ``gravitational mass and inertial mass are equal'' is a member of $K$---which would mean that it is not even a candidate \emph{explanandum}.  Kitcher assumes that $K$ is deductively closed and consistent.  Thus, if the statements of GR are members of $K$, then ``there is no such thing as gravitational mass'' is also a member of $K$, ruling out ``gravitational mass and inertial mass are equal''.  Indeed, on Kitcher's account of reduction, one would reduce Newtonian gravitation to GR by showing that the general argument patterns of Newtonian gravitation can be recovered from and extended as general argument patterns of GR.  But then one would not expect to be able to explain features of the world that can only be expressed in the Newtonian framework, because the very goal of the reduction would be to move one's explanatory arguments \emph{out} of the Newtonian framework and into GR.

Another worry comes from the opposite direction.  Even if $K$ were construed in such a way that the \emph{explanandum} were in $K$, it is still not clear that the explanation I have given would be included in the explanatory store.  This is because adding an additional argument of the sort I have given makes the explanatory store larger without explaining any new phenomena, since the argument patterns of GR are sufficient to explain the motion of bodies.  Thus adding such an explanation threatens to violate the minimality condition

A third worry is more general.  It is difficult to see what kind of argument pattern is being executed in the present example.  The thing to be explained is a very general observational feature of the world.  For this reason it is essentially singular.  One might be able to schematize the argument by reconstruing it as an argument concerning the inertial and gravitational masses for particular bodies, and then include a general argument pattern by which one shows that for any given body, the inertial and gravitational masses are identical.  But it is not clear that such explanations answer the original question.  Really we want to know why inertial and gravitational mass are \emph{always} the same, not why they happen to take the same value in any variety of situations.

These points suggest that the explanation I have given is an awkward fit with the unificationist account of explanation, just as with the DN account. Once again, this does not mean that the unificationist account cannot be adapted to fit the explanation I have given.  I think it probably can.  Indeed, it is hard to prove that a given explanation, if successful, is not a member of the maximally unified set of arguments over a set of beliefs.  My point is rather that to make the present explanation fit with the details of Kitcher's account, some adaptation of the unificationist position is likely necessary.  More importantly, even if one can find a modified unificationist account that would fit more naturally with the present explanation, it is not clear that it would do justice to the explanation I have given.  In other words, it does not seem that the \emph{reason} that the explanation I have given is explanatory has anything to do with the fact that it is an instantiation of a general, stringent, and unified argument pattern.  Indeed, I have suggested that it is not an instantiation of a general argument pattern at all---it is a singular explanation of a general feature of the world, as expressed in the language of Newtonian theory.

If anything, attempting to adapt the unificationist picture to include explanations of the present sort threatens to gloss over the important features of the explanation.  I would rather say that very many explanations in science have the character that Kitcher describes: they consist of arguments that are explanatory by virtue of how they fit some phenomenon into a systematized body of knowledge.  I find Kitcher's analysis of his own examples convincing, and I think that a broad class of explanations in physics fit well with the unificationist picture of explanation.  Modifying the unificationist account, which gets so many interesting cases right, seems counterproductive.  What we have here is simply a different kind of explanation.

\section{Let a thousand flowers bloom}\label{conclusion}

So far, I have argued that the explanation given in section \ref{explanation} is neither a causal explanation nor any kind of statistical explanation.  It is, at least roughly, a deductive argument, though it cannot easily be made to fit the strict logical structure demanded by the DN model. Likewise, the explanation seems to be a poor fit with the unificationist model, at least as described by \citet{Kitcher}.  It should be clear from my remarks in section \ref{introduction} that I do not take these accounts' inability to deal with the present explanation as an argument against the accounts. I have simply presented a scientific explanation of a different sort (and there are many possible sorts).

The remaining work is to identify some of the features of the present explanation and to try to characterize what makes it explanatory.  I have identified the most important features along the way, but it seems worthwhile to tie them together now.  As we have seen, the present explanation is an argument (broadly construed), and the \emph{explanandum} is a feature of the world (observed in some regimes) as represented in the language of Newtonian physics.  The explanation involves a translation between GR and Newtonian gravitation that is accomplished via the two step limiting process described above. This translation is necessary because the question is asked in the language of one theory (indeed, it only makes sense in that theory), but it is of a form that \emph{a fortiori} cannot be answered without appealing to physics that goes beyond the theory within which it is asked.  The explanatory demand is to show how, given some superseding theory, a general fact as expressed within one theory is really necessary or to be expected, at least within the regime in which the old theory is successful.

What is doing the explanatory work here?  One way of viewing the explanation is as a derivation of the \emph{explanandum} as a general constraint on the models of Newtonian gravitation that can be understood as good approximations of a relativistic world.  In other words, the explanation consists in showing that if the world is to be understood as a model of GR, then any model of Newtonian gravitation that successfully approximates the world, even in limited regimes, must satisfy the condition that $m_\mathcal{I}=m_{\mathcal{G}}$.  Insofar as we accept GR, and insofar there are contexts in which one can understand gravitation as a force, at least approximately, the coupling of a body to the gravitational field is given by the body's inertial mass.  The explanatory work, then, is done by presenting the details of the relationship between the two theories.

At the very beginning of this paper, I gave two other examples of why questions that I claimed would be well answered by explanations of the sort I have given here.  Those questions, and myriad others that one often encounters in physics, also call for explanations of general features of the world, at least within some particular regime.  They, too, are expressed within the framework and language of a particular theory.  And yet, there is good reason to believe that in answering these questions, we will exhibit important shortcomings of the theories in which the questions are asked, much like GR reveals the ways in which Newtonian gravitation is unsatisfactory.  Since they are open questions, I cannot say with certainty what their answers will be.  But I do maintain that they are asked in the spirit of the question that I focus on in the body of the paper, and moreover, they would be well-answered by explanations that involve showing how the present theory arises via some limiting procedure from a new, superceding theory. For this reason, I think that physicists are regularly making explanatory demands on future theories that call for explanations ill-treated by the philosophical literature on scientific explanation.

Now that I have explored the details of the kind of explanation I have in mind, let me return to what I described in the introduction as the principal moral of this discussion.  I have argued that the essential piece of the explanation I have given, the part that does the explanatory work, comes out in the details of the relationship between the theory in which the question is asked and a new theory that supercedes it.  I have also argued that explanatory demands in the spirit of the explanation I have given, demands that would be met by explanations of the sort described here, are an important part of research in physics.  Indeed, the question that I focused on in this paper, concerning the relationship between inertial and gravitational mass, played such a role in several research projects at the beginning of the 20th century: Lorand E\"otv\"os and Einstein both took the unexplained equality of inertial and gravitational mass to be of utmost importance in the study of gravitation. I think these two points, taken together, reveal something important about the structure of inquiry in physics.

First, in asking a certain kind of question, physicists implicitly acknowledge that current theory is incomplete---indeed, the particular questions they ask can be taken as attempts to identify where they expect current theory to break down.  But this acknowledgement is counterbalanced by an expectation that the framework and concepts of the present theory offer sufficiently good descriptions of the world that one should expect the questions to make sense, and moreover to have answers, even after a new theory has been discovered.  This is not to say that the new theory will look anything like the old theory.  But one nonetheless expects the conceptual frameworks of current theories to have sufficient stability that the present questions will be recoverable from the new theory, just as one was able to recover a statement about gravitational mass in an appropriate limit from a theory that does not support a notion of gravitational mass.

It seems to me that the prevalence of the sort of explanatory demands just described poses a significant problem for a certain class of views, often associated with Kuhn, concerning the incommensurability of scientific theories.  If theory change involves demands for and offers of explanations of the sort I have described, then not only is translation between distinct theories with overlapping domains possible, it is an essential part of how physicists understand their own work.  But I will not develop this idea further.  Suffice it to say that explanations, and particularly demands for explanation, play many roles in the creation and evolution of physical theories.  Some of these explanations will be of a kind with the one I have treated here.  And as Geroch suggests in the above quoted passage, to say any more would be a mistake.

\appendix
\section{Technical Details of Answer to Question 3}\label{details}

In the body of the paper I offer an answer to the question, ``Why are inertial and gravitational mass equal in Newtonian physics?''  Here I offer the technical details of that argument, pulling together certain strands of the literature that are not usually treated together systematically.  I will provide only a brief, formal review of geometrized and covariant standard Newtonian theory; I will mostly take familiarity with GR for granted.  For a pedagogical treatment of these subjects, including proofs of the theorems stated here and an explanation of the ``abstract index notation'' I will use throughout, see \citet{MalamentGR}.  For more details on the sense in which geometrized Newtonian theory arises as a limit from GR, see \citet{Kunzle}, \citet{Ehlers}, and \citet{MalamentNG}.

\subsection{Preliminary definitions}

We begin by defining the geometrical structures we will work with.  A model of GR can be defined as follows.  \begin{defn}\singlespacing A \emph{relativistic spacetime} is an ordered pair $(M,g_{ab})$, where $M$ is a smooth,\footnote{Here and in what follows, it should be assumed that we are limiting attention to smooth (i.e. infinitely differentiable) curves, fields, manifolds, etc., whether stated explicitly or not.} connected, four-dimensional manifold and $g_{ab}$ is a smooth, non-degenerate semi-Riemannian metric on $M$ with Lorentz signature $(+,-,-,-)$.\end{defn}\doublespacing  In a relativistic spacetime, the metric defines a lightcone structure at every point as follows.  Given any point $p$ and any vector $\xi^a$ in the tangent space $M_p$, we say that $\xi^a$ is \emph{timelike} if $g_{ab}\xi^a\xi^b>0$, \emph{spacelike} if $g_{ab}\xi^a\xi^b<0$, and \emph{null} if $g_{ab}\xi^a\xi^b=0$.  The length of any vector $\xi^a$ at a point is given by $||\xi^a||=|g_{ab}\xi^a\xi^b|^{1/2}$.  A (smooth) curve is timelike (resp. spacelike or null) if its tangent vector is at every point of the curve. There is a unique torsion-free derivative operator, $\nabla$, satisfying the metric compatibility condition $\nabla_ag_{bc}=\mathbf{0}$; $\nabla$, in turn, determines the Riemann curvature tensor $R^{a}{}_{bcd}$.

Massive point particles are represented by their worldlines, which are smooth future-directed timelike curves parametrized by arc-length.  With every point particle, there is an associated \emph{four-momentum}, $P^a$, defined at every point of the particle's worldline, whose length is the (inertial) \emph{rest mass}.  For a point particle with non-zero mass $m_{\mathcal{I}}$,\footnote{Since keeping track of the distinction between inertial and gravitational mass is important for the ultimate moral of the present discussion, I will label masses as inertial even in the context of GR and geometrized Newtonian gravitation, were strictly speaking there can be no ambiguity.   I will use capitalized calligraphic symbols for subscripts indicating labels to distinguish them from subscripts indicating index (i.e. tensor) structure.} we can write $P^a=m_{\mathcal{I}}\xi^a$, where $\xi^a$ is the tangent vector field to the particle's worldline (called the particle's \emph{four-velocity}).  The acceleration of a particle's wordline, $\xi^n\nabla_n\xi^a$, is determined by the relation $F^a=m_{\mathcal{I}}\xi^n\nabla_n\xi^a$, where $F^a$ represents the external forces acting on the particle.  In the absence of forces, massive test point particles traverse timelike geodesics.   More generally, we can associate with any matter field a smooth symmetric field $T^{ab}$, called the energy-momentum tensor.  $T^{ab}$ can be thought to encode the four-momentum density of the matter field as determined by any future-directed timelike observer at a point: for all points $p\in M$ and all unit, future-directed timelike vectors at $p$, $\xi^a$, the four-momentum of a matter field at $p$ as determined by $\xi^a$ is $P^a=T^a_{\;\;b}\xi^b$.  The curvature of spacetime is related to the energy-momentum tensor by \emph{Einstein's equation},\begin{equation}\label{EE}
R_{ab}=8\pi(T_{ab}-\frac{1}{2} Tg_{ab}),\end{equation} where $T=T^a_{\;\;a}$ and $R_{ab}=R^n{}_{abn}$ is the Ricci tensor.

We can now proceed to define a parallel structure for classical theories.
\begin{defn}\singlespacing
A \emph{classical spacetime} is an ordered quadruple $(M, t_{ab}, h^{ab},\nabla)$, where  $M$ is a smooth, connected, four-dimensional manifold; $t_{ab}$ is a smooth symmetric field on $M$ of signature $(1,0,0,0)$; $h^{ab}$ is a smooth symmetric field on $M$ of signature $(0,1,1,1)$; and $\nabla$ is a derivative operator on $M$ compatible with $t_{ab}$ and $h^{ab}$, i.e. it satisfies $\nabla_a t_{bc}=\mathbf{0}$ and $\nabla_a h^{bc}=\mathbf{0}$. We additionally require that $t_{ab}$ and $h^{ab}$ are orthogonal, i.e. $t_{ab}h^{bc}=\mathbf{0}$.\end{defn}\doublespacing
Note that ``signature,'' here, has been extended to cover the degenerate case.  We can see immediately from the signatures of $t_{ab}$ and $h^{ab}$ that neither is invertible.  Hence in general neither $t_{ab}$ nor $h^{ab}$ can be used to raise and lower indices.

$t_{ab}$ can be thought of as a temporal metric on $M$ in the sense that given any vector $\xi^a$ in the tangent space at a point, $p$, $||\xi^a||=(t_{ab}\xi^a\xi^b)^{1/2}$ is the \emph{temporal length} of $\xi^a$ at that point.  If the temporal length of $\xi^a$ is positive, $\xi^a$ is \emph{timelike}; otherwise, it is \emph{spacelike}.  At any point, it is possible to find a covector $t_a$, unique up to sign, such that $t_{ab}=t_at_b$.  If there is a continuous, globally defined vector field $t_a$ such that at every point, $t_{ab}=t_at_b$, then the spacetime is \emph{temporally orientable} (we will encode the assumption that a spacetime is temporally oriented by replacing $t_{ab}$ with $t_a$ in our definitions of classical spacetimes).  $h^{ab}$, meanwhile, can be thought of as a spatial metric.  However, since there is no way to lower the indices of $h^{ab}$, we cannot calculate the spatial length of a vector directly.  Instead, we rely on the fact that if $\xi^a$ is a spacelike vector (as defined above), then there exists a covector $\sigma_a$ such that $\xi^a=h^{ab}\sigma_b$.  The \emph{spatial length} of $\xi^a$ can then be defined as $(h^{ab}\sigma_a\sigma_b)^{1/2}$.  (This spatial length is independent of the choice of $\sigma_a$; if $\xi^a$ is not a spacelike vector, then there is no way to assign it a spatial length.)  Note, too, that it is possible to define the Riemann curvature tensor $R^{a}_{\;\;bcd}$ and the Ricci tensor $R_{ab}$ with respect to $\nabla$ as in GR (or rather, as in differential geometry generally).  Flatness ($R^a_{\;\;bcd}=\mathbf{0}$) carries over intact from GR; we say a classical spacetime is \emph{spatially flat} if $R^{abcd}=R^a_{\;\;nmq}h^{bn}h^{cm}h^{dq}=\mathbf{0}$.  It turns out that this latter condition is equivalent to $R^{ab}=h^{an}h^{bm}R_{nm}=\mathbf{0}$.\footnote{See \citet[Prop. 4.15]{MalamentGR}.}

We describe matter in close analogy with GR.  Massive point particles are again represented by their worldlines, which are smooth future-directed timelike curves parameterized by elapsed time. For a point particle with (inertial) mass $m_{\mathcal{I}}$, we can always define a smooth unit vector field $\xi^a$ tangent to its worldline, again called the four-velocity, such that we can define a four-momentum field $P^a=m_{\mathcal{I}}\xi^a$.  The mass of the particle is now given by the temporal length of its four-momentum.  In similar analogy to the relativistic case, we can associate with any matter field a smooth symmetric field $T^{ab}$, now called the mass-momentum tensor.  $T^{ab}$ once again encodes the four-momentum density of the matter field as determined by a future directed timelike observer at a point, but in this case all observers agree on the four-momentum density at $p$: $P^a=t_bT^{ab}$.  Contracting once more with $t_b$ yields the mass density, $\rho=t_at_bT^{ab}$.

In the present covariant, four-dimensional language, standard Newtonian theory can be expressed as follows.  Let $(M,t_a,h^{ab},\nabla)$ be a classical spacetime.  We require that $\nabla$ is flat (i.e. $R^a_{\;\;bcd}=\mathbf{0}$).  We begin by considering the dynamics of a test point particle with inertial mass $m_{\mathcal{I}}$ and four-velocity $\xi^a$.  As in GR, the force on a particle is related to the acceleration of its worldine by $F^a=m_{\mathcal{I}}\xi^n \nabla_n\xi^a$.  In the absence of external forces, a massive test point particle undergoes geodesic motion.  If the total mass-momentum content of spacetime is described by $T^{ab}$, we require that the conservation condition holds, i.e. at every point $\nabla_aT^{ab}=\mathbf{0}$.  To add gravitation to the theory, we can represent the gravitational potential as a smooth scalar field $\varphi$ on $M$.  $\varphi$ is required to satisfy Poisson's equation, $\nabla_a\nabla^a\varphi=4\pi\rho$ (where $\nabla^a$ is shorthand for $h^{ab}\nabla_b$).  Gravitation is considered a force; in general, the gravitational force on a point particle is moderated by its gravitational mass, according to $F^a_{\mathcal{G}}=-m_{\mathcal{G}}\nabla^a\varphi$.

In geometrized Newtonian theory we again begin with a classical spacetime $(M,t_a,h^{ab},\nabla)$, but now we allow $\nabla$ to be curved.  The dynamics of a point particle with inertial mass $m_{\mathcal{I}}$ and four-velocity $\xi^a$ are again given by $F^a=m_{\mathcal{I}}\xi^n\nabla_n\xi^a$; likewise, free massive test point particles undergo geodesic motion.  However, the geodesics are now determined relative to $\nabla$, which is not necessarily flat.  The conservation condition is again expected to hold.  Gravitational interactions are now seen to be the result of the curvature of spacetime, which in turn is determined by a geometrized form of Poisson's equation,\begin{equation}\label{GeoPoisson}R_{ab}=4\pi\rho t_a t_b.\end{equation}  The geometrized Poisson's equation forces spacetime to be spatially flat, because if Poisson's equation holds, then $R^{ab}=4\pi\rho h^{an}h^{bm}  t_n t_m=\mathbf{0}$ by the orthogonality condition on the metrics.

\subsection{Relations between the theories}

We are particularly interested in the relationship between these three theories.  Several results are available.  First, it is always possible to ``geometrize'' a gravitational field on a flat classical spacetime---that is, we can always move from the covariant formulation of standard Newtonian gravitation to geometrized Newtonian gravitation, via a result due to Andrzej \citet{Trautman}.
\begin{prop}[Trautman Geometrization Lemma.]\label{geometrization}\singlespacing \emph{\citep[Slightly modified from][Prop. 4.2.1.]{MalamentGR}} Let $(M,t_a,h^{ab},\overset{f}{\nabla})$ be a flat classical spacetime.  Let $\varphi$ and $\rho$ be smooth scalar fields on $M$ satisfying Poisson's equation, $\overset{f}{\nabla}_a\overset{f}{\nabla}\,^a\varphi=4\pi\rho$.  Finally, let $\overset{g}{\nabla}=(\overset{f}{\nabla},C^a_{\;\;bc})$,\footnote{This notation is explained in \citet[Prop. 1.7.3]{MalamentGR}.  Briefly, if $\nabla$ is a derivative operator on $M$, then any other derivative operator on $M$ is determined relative to $\nabla$ by a smooth symmetric (in the lower indices) tensor field, $C^a_{\;\;bc}$, and so specifying the $C^{a}_{\;\;bc}$ field and $\nabla$ is sufficient to uniquely determine a new derivative operator.} with $C^a_{\;\;bc}=-t_bt_c\overset{f}{\nabla}\,^a\varphi$.  Then $(M,t_a,h^{ab},\overset{g}{\nabla})$ is a classical spacetime; $\overset{g}{\nabla}$ is the unique derivative operator on $M$ such that given any timelike curve with tangent vector field $\xi^a$, \begin{equation}\label{geoEquiv}\tag{G}\xi^n\overset{g}{\nabla}_n\xi^a=\mathbf{0}\Leftrightarrow \xi^n\overset{f}{\nabla}_n\xi^a=-\overset{f}{\nabla}\,^a\varphi;\end{equation} and the Riemann curvature tensor relative to $\overset{g}{\nabla}$, $\overset{g}{R}\,^a_{\;\;bcd}$, satisfies \begin{align}&\overset{g}{R}_{ab}=4\pi\rho t_a t_b\tag{CC1}\label{CC1}\\ &\overset{g}{R}{}^a_{\;\;b}{}^c_{\;\;d}=\overset{g}{R}{}^{c}_{\;\;d}{}^a_{\;\;b}\tag{CC2}\label{CC2}\\
&\overset{g}{R}{}^{ab}{}_{cd}=\mathbf{0}\tag{CC3}\label{CC3}.\end{align}\end{prop}\doublespacing

Trautmann showed that it is also possible to go in the other direction.  That is, given a curved classical spacetime, it is possible to recover a flat classical spacetime and a gravitational field, $\varphi$---so long as the curvature conditions \eqref{CC1}-\eqref{CC3} are met.  \begin{prop}[Trautman Recovery Theorem.]\label{recovery}\singlespacing \emph{\citep[Slightly modified from][Prop. 4.2.5.]{MalamentGR}} Let $(M, t_a,h^{ab},\overset{g}{\nabla})$ be a classical spacetime that satisfies eqs. \eqref{CC1}-\eqref{CC3} for some smooth scalar field $\rho$.  Then, at least locally on $M$, there exists a smooth scalar field $\varphi$ and a flat derivative operator on $M$, $\overset{f}{\nabla}$, such that $(M,t_a,h^{ab},\overset{f}{\nabla})$ is a classical spacetime; \eqref{geoEquiv} holds of any timelike curve; and $\varphi$ and $\overset{f}{\nabla}$ together satisfy Poisson's equation, $\overset{f}{\nabla}_a\overset{f}{\nabla}\,^a\varphi=4\pi\rho$.\end{prop} It is worth pointing out that the pair $(\overset{f}{\nabla},\varphi)$ is not unique.  It is also worth pointing out that whenever we begin with standard Newtonian theory and move to geometrized Newtonian theory, it is always possible to move back to the standard theory, because Prop. \ref{geometrization} guarantees that the curvature conditions \eqref{CC1}-\eqref{CC3} are satisfied.

We can now ask how either of these classical theories relate to GR.  The answer is that geometrized Newtonian theory arises as a limiting case of GR, for a properly constructed limit.  (For full details of this limiting procedure, see \citet[Sec. 5]{MalamentNG}.)  Intuitively, we will begin with a relativistic spacetime, and then allow the lightcone structure at every point to open in such a way that, in the limit, the lightcones at every point become degenerate.  Since the lightcone structure in a sense determines the speed of light, allowing the lightcones to widen in this fashion captures a sense in which one might allow the speed of light to go to infinity.

The limit can be constructed as follows. Consider a manifold $M$ admitting a one-parameter family of nondegenerate Lorentz metrics $g_{ab}(\lambda)$ (where $\lambda$ ranges over some interval $(0,k)\subseteq\mathbb{R}$) that satisfy two conditions:\renewcommand{\theenumi}{(Lim\arabic{enumi})}
\renewcommand{\labelenumi}{\theenumi}
\begin{enumerate}
\item\label{Lim1} $\lim_{\lambda\rightarrow 0}g_{ab}(\lambda)=t_{a}t_{b}$ for some non-vanishing closed field $t_a$;\footnote{If $t_a$ is a non-vanishing closed field, the product $t_{ab}=t_at_b$ automatically has signature $(1,0,0,0)$.}
\item\label{Lim2} $\lim_{\lambda\rightarrow 0}\lambda g^{ab}(\lambda)=-h^{ab}$ for some field $h^{ab}$ of signature $(0,1,1,1)$.
\end{enumerate}
For any $\lambda\in(0,k)$, we can associate with $g_{ab}(\lambda)$ the unique covariant derivative operator compatible with $g_{ab}(\lambda)$, $\overset{\lambda}{\nabla}$, as well as the Ricci curvature tensor associated with $\overset{\lambda}{\nabla}$, $\overset{\lambda}{R}_{ab}$.  Thus the one-parameter family of metrics generates a one-parameter
family of compatible derivative operators and curvature tensors.  Suppose further that, for any $\lambda\in (0,k)$, we can define a smooth symmetric field $T^{ab}(\lambda)$ that together with $g_{ab}(\lambda)$ and its associated Ricci tensor satisfy:
\begin{enumerate}\setcounter{enumi}{2}
\item\label{Lim3} For all $\lambda\in(0,k)$, $\overset{\lambda}{R}_{ab}=8\pi\left(T_{ab}(\lambda)-\frac{1}{2}g_{ab}(\lambda)T(\lambda)\right)$, where $T(\lambda)=T_{ab}(\lambda)g^{ab}(\lambda)$; and
\item\label{Lim4} $\lim_{\lambda\rightarrow 0}T^{ab}(\lambda)=T^{ab}$ for some smooth symmetric field $T^{ab}$ on $M$.
\end{enumerate}\renewcommand{\theenumi}{\arabic{enumi}.}
\renewcommand{\labelenumi}{\theenumi}

When these conditions hold, it is possible to prove the following result.
\begin{prop}[Classical Limit of GR.]\label{limit}\singlespacing \emph{\citep[Adapted from][Props. on Limits 1 \& 2]{MalamentNG}} Fix a smooth, connected, four-dimensional manifold $M$ and assume $\lambda$ is a real-valued variable taking all values on an interval $(0,k)$.  Suppose that for each $\lambda$ on an interval $(0,k)$, there exist smooth symmetric fields $g_{ab}(\lambda)$ and $T_{ab}(\lambda)$ on $M$ such that $(M,g_{ab}(\lambda))$ is a relativistic spacetime and for each $\lambda$, $g_{ab}(\lambda)$ and $T_{ab}(\lambda)$ collectively satisfy conditions \ref{Lim1}-\ref{Lim4}. Then there exists a derivative operator $\nabla_a$ on $M$ such that $\lim_{\lambda\rightarrow 0}\overset{\lambda}{\nabla}_a=\nabla_a$,\,\footnote{What does it mean for a sequence of derivative operators to converge?  Suppose that $\tilde{\nabla}_a$ is a fixed auxiliary derivative operator on $M$.  Then for each $\overset{\lambda}{\nabla}_a$, there is a smooth symmetric field $C^a_{\;\;bc}(\lambda)$ such that $\overset{\lambda}{\nabla}_a=(\tilde{\nabla}_a,C^{a}_{\;\;bc}(\lambda))$.  Now suppose that there is another derivative operator on $M$, $\nabla_a=(\tilde{\nabla}_a,C^{a}_{\;\;bc})$.  We can say that $\lim_{\lambda\rightarrow 0}\overset{\lambda}{\nabla}_a=\nabla_a$ if $\lim_{\lambda\rightarrow 0}C^a_{\;\;bc}(\lambda)=C^a_{\;\;bc}$.} and for which $(M,t_{a},h^{ab},\nabla_a)$ is a classical spacetime satisfying $R^{a}_{\;\;b}{}^c_{\;\;d}=R^c_{\;\;d}{}^a_{\;\;b}$.  Moreover, there exists a smooth field $\rho$ on $M$ such that $\lim_{\lambda\rightarrow 0} T_{ab}(\lambda)=\rho t_{a}t_{b}$, which satisfies $R_{ab}=4\pi\rho t_{a}t_{b}$.
\end{prop}
Prop. \ref{limit} gives the precise sense in which geometrized Newtonian gravitation is a limiting case of GR.

\subsection{Gravitational Mass in Newtonian Theory}

We have now done sufficient groundwork to offer a technically precise formulation of the explanation given in the body of the paper.  The argument was that by beginning with GR and then moving in the limit to standard Newtonian theory, one finds that a massive point particle's coupling to the gravitational field is given by its inertial mass.  This limit proceeds in two steps.  First, using Prop. \ref{limit}, one shows that geometrized Newtonian gravitation is a limiting case of GR.  Prop. \ref{recovery}, meanwhile, shows that when three curvature conditions, \eqref{CC1}-\eqref{CC3}, are satisfied, we can recover (covariant) standard Newtonian theory from geometrized Newtonian theory.  It is in the course of executing this two step process that one is forced to associate gravitational and inertial mass.

There is an important subtlety here.  To connect Props. \ref{recovery} and \ref{limit} and show that we can recover standard Newtonian theory as a limit from GR, we need to show that the classical spacetime we reach in the limit from GR in fact meets the three curvature conditions necessary to recover the standard theory. Prop. \ref{limit} gives that two of the curvature conditions, \eqref{CC1} and \eqref{CC2}, are satisfied automatically.  But what about \eqref{CC3}, $R^{ab}{}_{cd}=\mathbf{0}$?  In general, $R^{ab}{}_{cd}$ need \emph{not} vanish in a classical spacetime reached in the limit from GR.  It turns out that there is a more general recovery theorem, due to Hans-Peter \citet{Kunzle} and J\"urgen \citet{Ehlers}, that holds when $R^{ab}{}_{cd}\neq\mathbf{0}$.  But the theory that you recover in this case is not standard Newtonian gravitation---it is a non-geometrized generalization of standard Newtonian gravitational theory in which the gravitational potential field is replaced by a vector field and there is an additional contribution to the force law for a particle arising from a kind of universal rotation.  The third curvature condition is sufficient to guarantee that this rotational contribution vanishes and that the gravitational vector field can be written as the covariant derivative of a scalar potential.

For present purposes, establishing when $R^{ab}{}_{cd}=\mathbf{0}$ holds is unnecessary.  The important point is that condition \eqref{CC3} is both a necessary and sufficient condition for recovery of standard Newtonian physics via Prop. \ref{recovery}.  This is not a problem for the explanation, \emph{per se}, since insofar as we can recover standard Newtonian theory at all, we can do so only from families of relativistic spacetimes that converge to classical spacetimes satisfying \eqref{CC3}.

I can now state the precise claim, the proof of which amount to the formal explanation.
\begin{prop}\singlespacing Let $(M, t_a, h^{ab},\overset{f}{\nabla})$ be a flat classical spacetime and let $\varphi$ be a gravitational field defined on that spacetime as in standard Newtonian gravitation.  Suppose further that $(M, t_a, h^{ab},\overset{f}{\nabla})$ and $\varphi$ arise via the two-step limiting process just described (which is only possible if the intermediate curved classical spacetime satisfies \eqref{CC3}).  Consider a massive point particle with inertial mass $m_{\mathcal{I}}$ traversing a timelike curve in $M$, $\gamma$, with tangent vector field $\xi^a$, under the influence of only gravitational force.  Then the gravitational force experienced by the massive point particle is\begin{equation}F^a_{\mathcal{G}}=m_{\mathcal{I}}\xi^n\overset{f}{\nabla}_n\xi^a=-m_{\mathcal{I}}\overset{f}{\nabla}{}^a\varphi.\end{equation}  In other words, the particle's gravitational mass is equal to its inertial mass.\end{prop}
Proof.  By assumption, $(M, t_a, h^{ab},\overset{f}{\nabla})$ and $\varphi$ arise via the two-step limiting process described above.  Thus there exists a (curved) classical spacetime $(M, t_a, h^{ab},\overset{g}{\nabla})$ satisfying \eqref{CC3} from which $(M,t_a,h^{ab},\overset{f}{\nabla})$ can be recovered.  Since the particle experiences no non-gravitational force, we know from the geodesic principle of geometrized Newtonian gravitation that $\gamma$ must be a geodesic relative to $\overset{g}{\nabla}$.  Meanwhile, by Prop. \ref{recovery}, we know that if $\gamma$ is a geodesic relative to $\overset{g}{\nabla}$, then $\xi^n\overset{f}{\nabla}_n\xi^a=-\overset{f}{\nabla}{}^a\varphi$.  Thus we have the acceleration of the particle's worldline, which we can plug into $F^a=m_{\mathcal{I}}\xi^n\overset{f}{\nabla}_n\xi^a$ to find the gravitational force on the particle.  We see that $F^a_{\mathcal{G}}=m_{\mathcal{I}}\xi^n\overset{f}{\nabla}_n\xi^a=-m_{\mathcal{I}}\overset{f}{\nabla}{}^a\varphi$, as required.  It follows that the particle's coupling to the gravitational field is given by its inertial mass.\hspace{.25in}$\square$

\singlespacing
\bibliography{explanation}
\bibliographystyle{mla}
\end{document}